\documentclass[letterpaper]{article}

\usepackage[T1]{fontenc}

\usepackage{geometry}
\usepackage{setspace}
\usepackage{ulem}

\usepackage[dvipdfmx]{graphicx}
\usepackage{here}
\usepackage{textcomp}
\usepackage{amsmath,amssymb}
\newfloat{scheme}{htbp}{los}
\floatname{scheme}{Scheme}
\floatname{chart}{Chart}
\newfloat{graph}{htbp}{loh}

\usepackage{chemformula} 
\usepackage[version = 4]{mhchem} 

\usepackage{authblk}
\author[1]{Kazumasa Iida*}
\affil[1]{College of Industrial Technology, Nihon University, Narashino, Japan}
\author[2]{Kai Walter}
\affil[2]{Institute for Technical Physics, Karlsruhe Institute of Technology, Eggenstein-Leopoldshafen, Germany}
\author[3]{Takafumi Hatano}
\affil[3]{Department of Materials Physics, Nagoya University, Nagoya, Japan}
\author[1]{Kose Morinaga}
\author[2]{Manuela Erbe}
\author[4]{Hongye Gao}
\affil[4]{The Ultramicroscopy Research Center, Kyushu University, Fukuoka, Japan}
\author[4, 5, 6]{Satoshi Hata}
\affil[5]{Department of Advanced Materials Science and Engineering, Kyushu University, Kasuga, Fukuoka, Japan}
\affil[6]{Interdisciplinary Graduate School of Engineering Sciences, Kyushu University, Kasuga, Fukuoka, Japan}
\author[2]{Jens H\"{a}nisch}

\title{Freestanding GdBa$_2$Cu$_3$O$_{7-\delta}$ Thin Films via Optimized Buffer Layer Design: Preserving Superconducting Properties}
\date{*Email: iida.kazumasa@nihon-u.ac.jp}

\begin{document}

\maketitle

\begin{abstract}
 Freestanding GdBa$_2$Cu$_3$O$_{7-\delta}$ (GdBCO) superconducting thin films were fabricated using a water-soluble Sr$_3$Al$_2$O$_6$ (SAO) sacrificial layer in combination with thermal release tape. An amorphous Al$_2$O$_3$ capping layer was introduced to suppress crack formation during the lift-off process. The influence of buffer-layer design inserted between the GdBCO and SAO layers was systematically investigated with respect to structural integrity and superconducting properties after lift-off. A LaAlO$_3$/SrTiO$_3$ bilayer buffer was found to be essential for maintaining epitaxial growth and a superconducting transition temperature ($T_\mathrm{c}$) of approximately 92\,K after lift-off, comparable to that of the as-grown films. In contrast, a reversed SrTiO$_3$/LaAlO$_3$ bilayer and single-layer buffer structures led to a suppression of $T_\mathrm{c}$, highlighting the critical role of stacking sequence. These results demonstrate that optimization of the buffer-layer design is a key factor for realizing high-quality freestanding GdBCO films while maintaining their superconducting characteristics.
\end{abstract}

\section*{Keywords}
free-standing membrane, cuprate superconductors, buffer-layer design, oxide heterostructures, superconducting properties
\section{1. INTRODUCTION}

Freestanding thin films offer a powerful means to evaluate the intrinsic properties of materials by eliminating substrate-induced effects. Free from substrate-induced strain and lattice mismatch, these systems enable the extraction of intrinsic structural and electronic characteristics, providing critical insights into anisotropy and the fundamental aspects of their physical properties. Such an approach allows property evaluations that are difficult to achieve with conventional epitaxial films and plays an important role in both the exploration of new materials and the validation of theoretical predictions.

From a device application standpoint, most modern semiconductor and electronic devices are based on silicon technology, creating strong demand for techniques that enable the stacking and integration of diverse functional thin films on Si substrates. However, the range of materials that can be directly grown on Si is limited, which has restricted broader device applications. Freestanding thin films have therefore attracted considerable attention as an effective strategy to overcome this limitation. 
Furthermore, functional freestanding films can be transferred not only onto Si but also onto arbitrary substrates \cite{ref2, ref3}. Owing to this versatility, freestanding thin films are also regarded as highly promising for applications in wearable devices and flexible electronics \cite{ref4, ref5}.

Research on freestanding thin films has advanced since the introduction of water-soluble sacrificial layers such as Sr$_3$Al$_2$O$_6$ (SAO) \cite{ref6} and BaO \cite{ref7, ref8}, and their application has now been extended to a wide range of oxides. For cuprate superconductors, freestanding films of YBa$_2$Cu$_3$O$_{7-\delta}$ (YBCO) \cite{ref9, ref10}, GdBa$_2$Cu$_3$O$_{7-\delta}$ (GdBCO)  \cite{ref11}, and Pr$_{1.85}$Ce$_{0.15}$CuO$_{4-\delta}$  \cite{ref12} have been reported. More recently, Sr$_4$Al$_2$O$_7$ has been proposed as a new water-soluble sacrificial layer  \cite{ref13}, and this approach has also enabled the fabrication of freestanding films of nickelate superconductors  \cite{ref14}.

However, YBCO superconductors are chemically unstable in water \cite{ref15}, making the straightforward removal of sacrificial layers with deionized water difficult. Consequently, etching in alkaline solutions with pH$\approx$12 has been employed \cite{ref9, ref10}. In contrast, there is only one report of  freestanding REBCO films obtained using deionized water, in that case GdBCO \cite{ref11}. When REBCO (RE: rare earth elements) is grown directly on SAO, the superconducting transition temperature ($T_\mathrm{c}$) is reduced, and X-ray diffraction (XRD) measurements indicate that the resulting films lack epitaxial growth. Moreover, at the high growth temperatures ($>$730$^\circ$C) required for REBCO deposition, interdiffusion between SAO and REBCO may occur, potentially leading to the critical issue that SAO no longer dissolves in water. Indeed, in LaMnO$_3$/SAO systems, interdiffusion has been reported to inhibit the dissolution of SAO \cite{ref16}. To mitigate this effect, a thin intermediate SrTiO$_3$ (STO) layer inserted between LaMnO$_3$ and SAO has been demonstrated to effectively suppress interdiffusion.

Following this approach, the insertion of a LaAlO$_3$ (LAO)/STO bilayer between YBCO and SAO has been reported to enable the fabrication of freestanding films without compromising $T_\mathrm{c}$ \cite{ref9}. However, no magnetization measurements were carried out, leaving it uncertain whether bulk superconducting properties were preserved. In this study, we aimed to fabricate GdBCO freestanding films that retain bulk superconducting properties by introducing various intermediate layers. As a result, a single intermediate layer between the SAO and GdBCO layers was inadequate. Only the LaAlO$_3$/SrTiO$_3$ bilayer configuration successfully maintained the superconducting properties, whereas the reversed SrTiO$_3$/LaAlO$_3$ sequence failed to do so, clearly evidencing the critical role of buffer-layer sequence. After removal of the sacrificial layer with deionized water, magnetization measurements indicated that superconducting properties were well preserved after lift-off, with $T_\mathrm{c}$$\approx$92\,K.

\section{2. EXPERIMENTAL SECTION}
All films were fabricated by pulsed laser deposition (PLD) using a KrF excimer laser with a wavelength of 248\,nm. STO(001) and LAO(001) substrates with dimensions of 10\,mm$\times$10\,mm and a thickness of 0.5\,mm were employed. The substrates were cut into 5\,mm$\times$5\,mm pieces, and STO substrates were surface-treated according to Ref.\,\cite{ref17}. The atomic force microscope (AFM) measurements confirmed that the treated STO substrate has an atomically flat surface (Supplementary Information, Fig.\,S1). The substrate was then fixed to a holder with Ag paste and introduced into the deposition chamber.

SAO was deposited under conditions of oxygen pressure 1.3\,Pa, substrate temperature 700$^\circ$C, laser repetition rate $f$=3\,Hz, and energy density $\epsilon$=1.5\,J/cm$^2$. Subsequently, STO and LAO were deposited as intermediate layers under the same conditions as SAO. GdBCO was then deposited under conditions of oxygen pressure 40\,Pa, substrate temperature 775$^\circ$C, $f$=10\,Hz, and $\epsilon$=1.5\,J/cm$^2$. The nominal thicknesses of the SAO, STO, LAO, and GdBCO layers were set to 45\,nm, 35\,nm, 20\,nm, and 175\,nm, respectively. After deposition, the samples were removed from the chamber and subjected to post-annealing in a tube furnace at 400$^\circ$C for 3\,h in 1\,atm of O$_2$. Table\,\ref{tab:samples} summarizes the films fabricated in this study.

\begin{table}[t]
\centering
\caption{Sample names and buffer-layer stacking sequences, including reference GdBCO films grown on single-crystal substrates.}
\begin{tabular}{cc}
\hline
Sample name           &  Film stacking structure   \\  \hline
Film A              		&   GdBCO/LAO/STO/SAO/STO-sub.                          \\
Film B                       &   GdBCO/STO/LAO/SAO/STO-sub.                          \\
Film C                       &   GdBCO/LAO/SAO/STO-sub.                                  \\
Film D                       &   GdBCO/STO/SAO/STO-sub.                                  \\
STD \#1                   &  GdBCO/STO-sub.                                                     \\ 
STD \#2                   &  GdBCO/LAO-sub.                                                      \\ \hline
\end{tabular}
\label{tab:samples}
\end{table}

Lift-off experiments were carried out for Film A as follows. To suppress crack formation during lift-off, an amorphous Al$_2$O$_3$ (a-Al$_2$O$_3$) capping layer was first deposited on the film surface by PLD at room temperature \cite{ref18}. A thermal release tape (Somar Co., Ltd., Somatac PS-2071TE; polyethylene terephthalate (PET) support layer 100\,$\mu$m) 
was then attached to the surface of the film, and the sample was immersed in deionized water. After approximately 2\,h, the film was successfully separated from the substrate.

The phase purities and the crystalline structure of the films were characterized by XRD with Cu K$\alpha$ radiation (SmartLab and UltimaIV, Rigaku, and D8 Discover, Bruker). 

Cross-sectional scanning transmission electron microscopy (STEM) specimens were prepared using a lift-out technique with a dual-beam focused ion beam (FIB) system (Scios, Thermo Fisher Scientific Inc.). Microstructural analysis was performed by STEM using a Thermo Fisher Scientific Titan G2 Cubed 60--300 microscope equipped with bright-field (BF) and annular dark-field (ADF) detectors. Chemical compositional analysis was carried out by energy-dispersive X-ray spectroscopy (EDS). All measurements were conducted at an acceleration voltage of 300\,kV.

The superconducting transition temperature $T_\mathrm{c}$ of as-grown films was measured using a four-probe method in a physical property measurement system (PPMS, Quantum Design, Inc.). The temperature dependence of magnetization for the lift-off samples was evaluated with a magnetic property measurement system (MPMS, Quantum Design, Inc.) under zero-field-cooled (ZFC) and field-cooled (FC) conditions. For the ZFC measurement, the samples were cooled to 35\,K in zero magnetic field, after which a magnetic field of 4.7\,Oe was applied parallel to the crystallographic $ab$--plane and the magnetization was recorded during warming. Subsequently, the FC magnetization was measured upon cooling under the same magnetic field.

\section{3. RESULTS AND DISCUSSION}
\subsection{3.1 Buffer-layer-dependent structural and transport properties before lift-off}

\begin{figure}[H]
\centering
	\includegraphics[width=11.5cm]{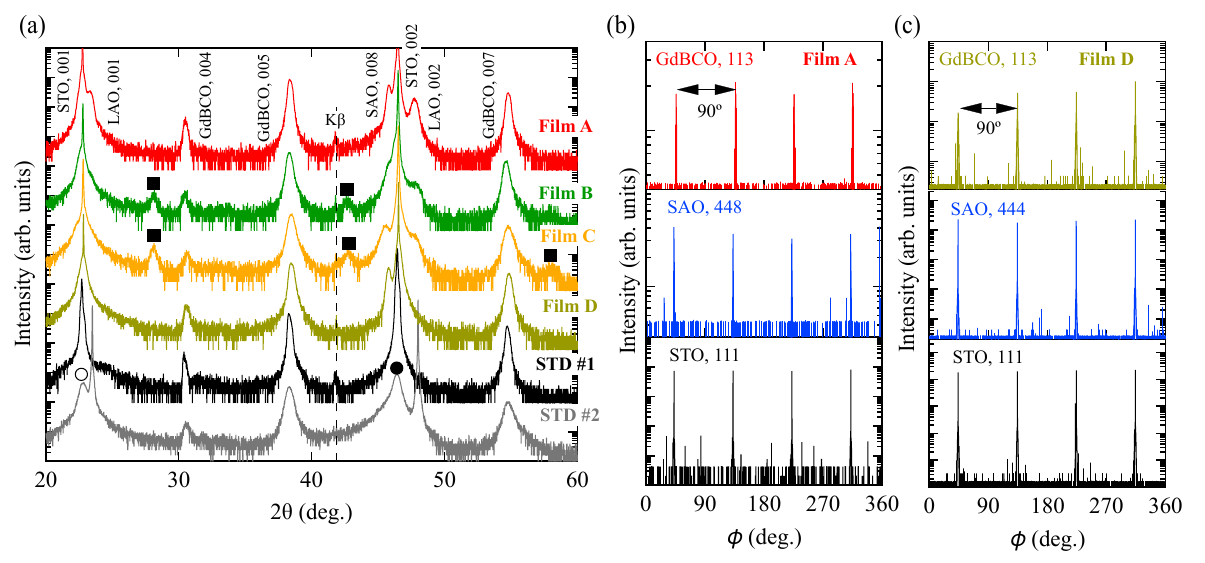}
	\caption{(a) XRD $2\theta/\omega$--scans of GdBCO films grown on various buffer layers with different stacking sequences. 
	$\blacksquare$ indicates SrLaAlO$_4$ (00$l$, $l$=2, 4, and 6). $\circ$ and \textbullet \,correspond to the 003 and 006 reflections of GdBCO. (b) The 113 $\phi$--scans of GdBCO, the 448 $\phi$--scans of SAO, and the 111 $\phi$--scans of STO substrate for Film A. (c) The 113 $\phi$--scans of GdBCO, the 444 $\phi$--scans of SAO, and the 111 $\phi$-scans of STO substrate for Film D.} \label{fig:Figure-1}
\end{figure}

The XRD $2\theta/\omega$--scans of the films listed in Table \,\ref{tab:samples} reveal an impurity phase with $c$-axis texture in Films B and C, which was identified as SrLaAlO$_4$ (00$l$, $l$=2, 4, and 6) [Fig.\,\ref{fig:Figure-1}(a)].

Possible interfacial reactions between LAO and SAO are suggested by the significantly weaker and less well-defined 008 reflection of SAO and 002 reflection of LAO in Films B and C compared with those in Film A. These findings are consistent with the ADF-STEM observations and the elemental analysis discussed below.

In contrast, Films A and D appear to be phase-pure within the detection limit of XRD (approximately 5\%) and all constituent layers exhibit 00$l$ orientation. Additionally, the in-plane orientation of both GdBCO and SAO was confirmed [Figs.\,\ref{fig:Figure-1}(b--c)], suggesting that the inserted layers LAO/STO in Film A and STO in Film D are also likely to be epitaxially grown.

\begin{figure}[t]
\centering
	\includegraphics[width=14cm]{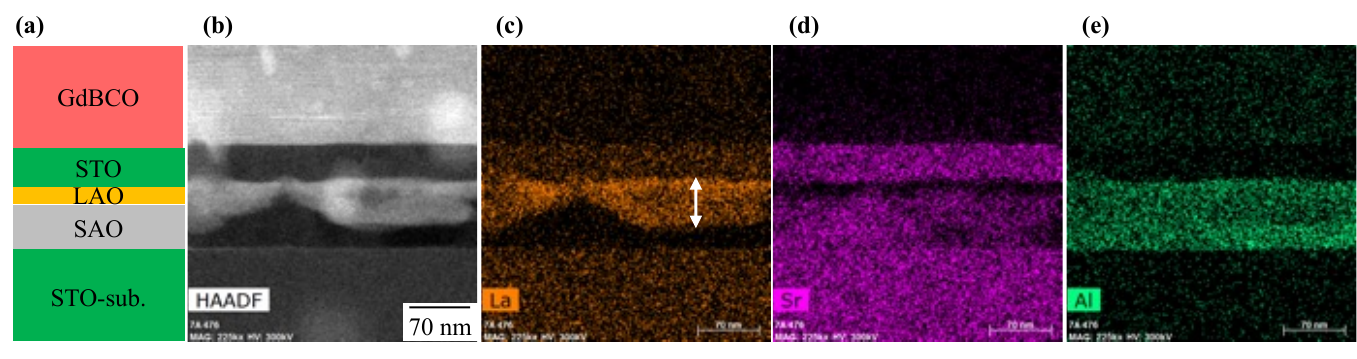}
	\caption{(a) Schematic illustration of the multilayer structure of Film B consisting of GdBCO/STO/LAO/SAO grown on an STO substrate.
(b) Cross-sectional ADF-STEM image of Film B. (c--e) Corresponding elemental maps obtained by EDS for La, Sr, and Al, respectively. The La-rich region extends to approximately 70\,nm, as indicated by the double arrows in (c).} \label{fig:Figure-2}
\end{figure}

The $c$-axis lattice parameters of Film A, Film D, STD \#1 (GdBCO grown on STO substrate), and \#2 (GdBCO grown on LAO substrate) are 11.714$\pm$0.012\,\AA, 11.692$\pm$0.018\,\AA, 11.710$\pm$0.007\,\AA, and 11.699$\pm$0.031\,\AA, respectively. All values agree with the bulk GdBCO lattice parameter ($c$=11.706\,\AA\,\cite{ref19}) within experimental uncertainty.

Cross-sectional ADF-STEM image together with the corresponding elemental mapping analyses were performed for Film B, as summarized in Fig.\,\ref{fig:Figure-2}. Although the nominal thickness of the LAO layer is approximately 20\,nm, the La-rich region extends to a thickness of about 70\,nm, as indicated by the double arrows in Fig.\,\ref{fig:Figure-2}(c). In addition, Sr is detected in a region extending up to 10\,nm below the STO interlayer [Fig.\,\ref{fig:Figure-2}(d)]. These results suggest that the formation of SrLaAlO$_4$ may have occurred in this region. A similar interfacial reaction may also be present in Film C.

\begin{figure}[H]
\centering
	\includegraphics[width=9.2cm]{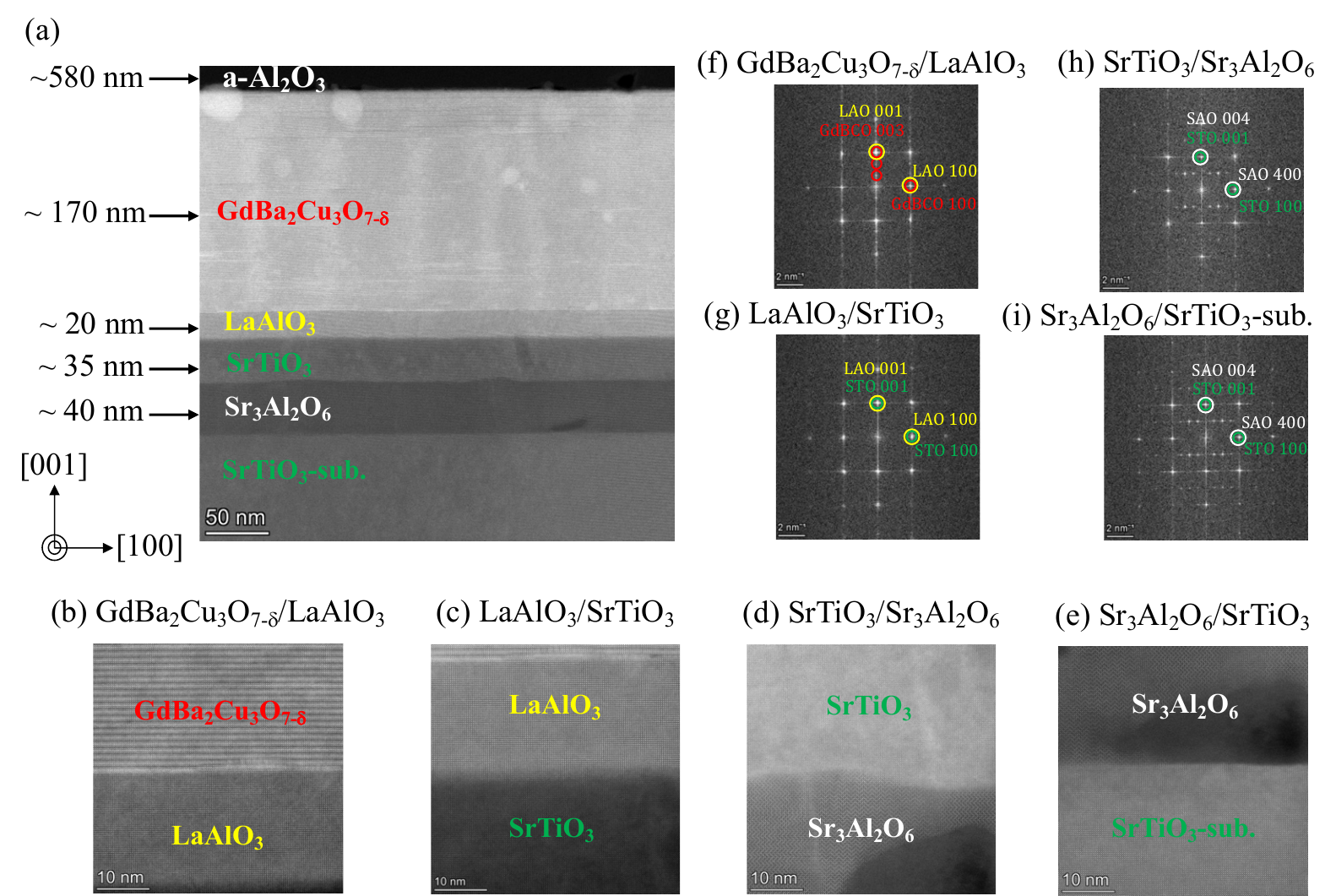}
	\caption{(a)The cross-sectional ADF-STEM image of Film A. Enlarged view of the interface between (b) GdBCO and LAO, (c) LAO and STO, (d) STO and SAO, and (e) SAO and STO substrates. The respective Fourier power spectra of the ADF-STEM images for (f) GdBCO and LAO, (g) LAO and STO, (h) STO and SAO, and (i) SAO and STO substrates.} \label{fig:Figure-3}
\end{figure}

The ADF-STEM microstructural analysis reveals that the GdBCO layer in Film A contains Gd$_2$O$_3$ precipitates, appearing as bright spherical features, as well as stacking faults [Fig.\,\ref{fig:Figure-3}(a)]. The interfaces between GdBCO and LAO, LAO and STO, STO and SAO, as well as SAO and the STO substrate are sharp in the ADF-STEM images, and no reaction layers are observed within the resolution of the present measurements [Figs.\,\ref{fig:Figure-3}(b--e)]. The corresponding Fourier power spectra of the ADF-STEM images indicate epitaxial alignment of each layer in a cube-on-cube configuration [Figs.\,\ref{fig:Figure-3}(f--i)].

\begin{figure}[t]
\centering
	\includegraphics[width=10cm]{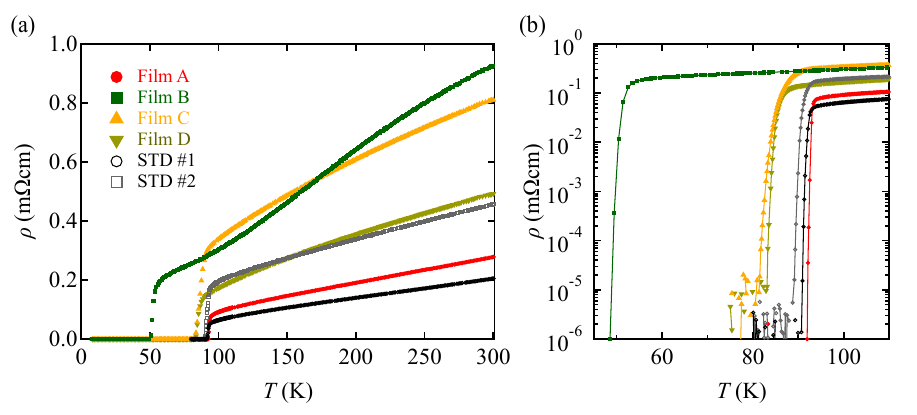}
	\caption{The temperature dependence of the resistivity for the films listed in Table\,\ref{tab:samples}: (a) over a wide temperature range and (b) semilogarithmic plots over a narrow temperature range.} \label{fig:Figure-4}
\end{figure}

The temperature dependence of the resistivity demonstrates that only Films A and STD \#1 exhibit zero resistivity ($T_\mathrm{c, 0}$) above 90\,K [Figs.\,\ref{fig:Figure-4}(a--b)]. Additionally, the normal-state resistivity remains below 0.28\,m$\Omega$cm. In contrast, the presence of the SrLaAlO$_4$ impurity phase (Films B and C) leads to a reduced $T_\mathrm{c, 0}$ and increased resistivity in the normal state. Notably, Film B exhibits a $T_\mathrm{c, 0}$ of only approximately 50\,K and the highest resistivity above 170\,K. Moreover, the shape of the normal-state resistivity is characteristic of an underdoped sample\,\cite{ref20}. The exponent $n$ in $\rho=\rho_0+AT^n$ ($\rho=\rho_0$ is the residual resistivity and $A$ is a constant) is approximately 1.6 for Film B, whereas it is close to 1 for other films (Supplementary Information, Fig.\,S2), which is associated with non-Fermi-liquid-like behavior.

Although Film D is phase-pure within the detection limit of XRD and epitaxially grown, its $T_\mathrm{c, 0}$ remains limited to approximately 83\,K. The origin of this behavior is not yet understood; however, a slight substitution of Ba by Sr cannot be excluded. A similar possibility may also be relevant for Films B and C. Notably, even a small degree of Sr substitution is known to significantly suppress $T_\mathrm{c}$. For example, a reduction rate of approximately 7.7\,K/$x$ has been reported for GdBa$_{2-x}$Sr$_x$Cu$_3$O$_{7-\delta}$ (Supplementary Information, Fig.\,S3) \cite{ref21}. Nevertheless, these results indicate that neither STO nor LAO alone is sufficient as an intermediate layer between GdBCO and SAO, and that a bilayer structure is required. Among the structures examined, the GdBCO/LAO/STO/SAO stacking sequence yields the most favorable superconducting properties.

\subsection{3.2 Preservation of bulk superconductivity after lift-off}
\begin{figure}[H]
\centering
	\includegraphics[width=9cm]{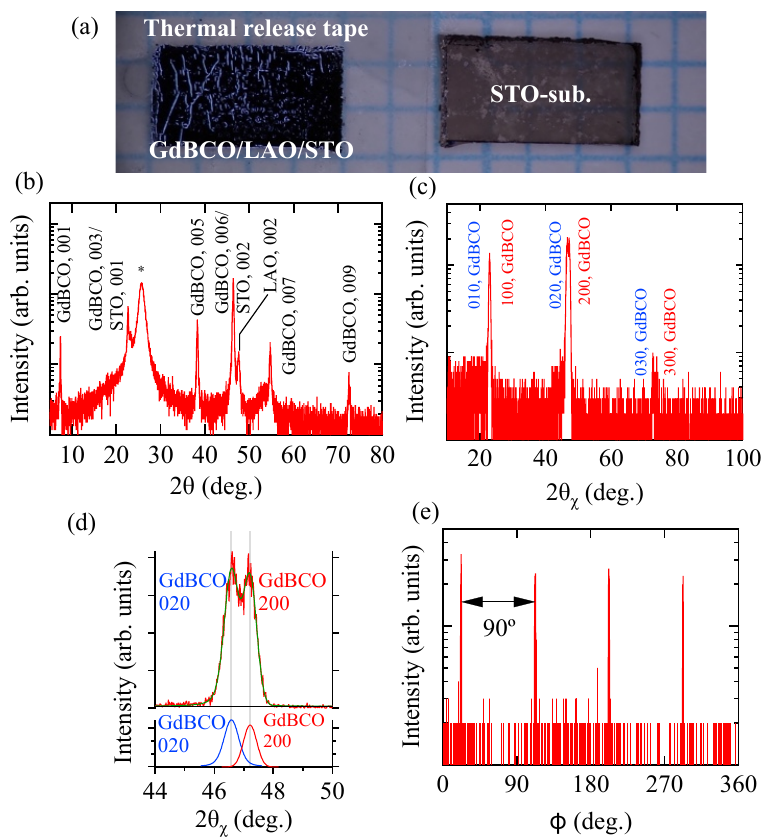}
	\caption{(a) Optical microscope image of a GdBCO/LAO/STO membrane on thermal release tape after lift-off, together with the STO substrate. Each grid square corresponds to 1\,mm$\times$1\,mm. (b) XRD $2\theta/\omega$--scan and (c) XRD $2\theta_\chi/\phi$--scan of the GdBCO/LAO/STO membrane shown in (a). The asterisk ($\ast$) corresponds to the XRD peak from the PET support layer. (d) Enlarged view of (c) in the range 44$^\circ\leq2\theta_\chi\leq50^\circ$, where the peak is deconvoluted into the 020 and 200 reflections of GdBCO. (e) Corresponding 020 $\phi$--scans of GdBCO.} \label{fig:Figure-5}
\end{figure}

A millimeter-scale freestanding GdBCO/LAO/STO membrane is successfully obtained after the lift-off process of Film A. Surface wrinkles are occasionally observed, whereas no macroscopic cracks are evident within the observed area [Fig.\,\ref{fig:Figure-5}(a)]. After lift-off, the surface of the GdBCO/LAO/STO membrane attached to a thermal release tape was brought into contact with a Si(001) substrate and heated to 120$^\circ$C in an attempt to transfer the film; however, the membrane fractured into multiple pieces. As a result, no successful transfer was achieved, and the GdBCO/LAO/STO membrane on the thermal release tape was instead subjected to structural and magnetic characterization.

The diffraction peaks of GdBCO/LAO/STO membrane, except for a broad peak at around 2$\theta\approx25^\circ$ originating from the PET support layer, are indexed to GdBCO and LAO, demonstrating that 00$l$ orientation is preserved [Fig.\,\ref{fig:Figure-5}(b)]. The 00$l$ reflections from STO overlap with the 00$l$ reflections of GdBCO, and therefore cannot be distinguished. The $c$-axis lattice parameter is evaluated to be 11.714\,\AA, which is identical to that of the as-grown Film A. 

The XRD $2\theta_\chi/\phi$--scan reveals a diffraction peak at around $2\theta\approx46^\circ$, corresponding to overlapping 200 and 020 reflections of GdBCO [Fig. 5(c--d)]. The in-plane lattice parameters $a$ and $b$ estimated from the 200 and 020 reflections are 3.85,\AA\, and 3.90\,\AA, respectively, which are close to those of the bulk crystal ($a$=3.899\,\AA\, and $b$=3.839\,\AA\,\cite{ref19}; the present work uses the standard orthorhombic convention ($a,b$), whereas ref.,\cite{ref19} uses reversed axis labeling). The corresponding 020 $\phi$--scan confirms fourfold symmetry [Fig.\,\ref{fig:Figure-5}(e)]. These results are consistent with the crystalline structure and epitaxial orientation being largely preserved after lift-off from the STO substrate.

\begin{figure}[H]
\centering
	\includegraphics[width=5cm]{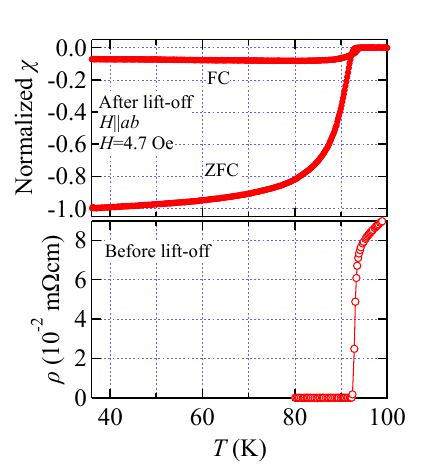}
	\caption{Temperature dependence of the normalized susceptibility of the GdBCO/LAO/STO membrane on thermal release tape. A small magnetic field of 4.7 \,Oe was applied parallel to the crystallographic $ab$--plane. The data were normalized to the value at 35\,K. For comparison, the temperature dependence of the resistivity of the corresponding film before lift-off is also shown.} \label{fig:Figure-6}
\end{figure}

The temperature dependence of the magnetization is shown in the upper panel of Fig.\,\ref{fig:Figure-6}, where the zero-field-cooled (ZFC) data are normalized to -1 at 35 K, assuming a fully superconducting volume. The zero-resistance temperature of the as-grown film (Fig.\,\ref{fig:Figure-6}, lower panel) coincides well with the onset $T_\mathrm{c}$ obtained from magnetization measurements after lift-off, with both showing a relatively sharp superconducting transition. The observed diamagnetic response indicates the formation of shielding currents within the film. This temperature marks the onset of a finite critical current and corresponds to the zero-resistance, or irreversibility, temperature in the resistivity-temperature dependence. Hence, the $T_\mathrm{c}$ values of the as-grown and lifted-off films are considered to be identical (see also Supplementary Information, Fig.\,S4), although the temperature dependence of the magnetization was not measured for the as-grown film.

Although the present study focuses on the superconducting properties of the freestanding membranes in their as-fabricated state, it is important to consider the effect of mechanical deformation such as bending. In general, superconducting properties in oxide thin films can be sensitive to strain and structural distortion. Therefore, bending-dependent measurements would provide further insight into the mechanical robustness of the superconducting state in freestanding GdBCO membranes. Such investigations are beyond the scope of the present study and will be addressed in future work.

The present results clearly demonstrate that the stacking sequence of the intermediate layers plays a critical role in governing both the structural integrity and superconducting properties of the freestanding GdBCO films. In particular, the LAO/STO bilayer configuration effectively preserves epitaxial growth and superconducting characteristics, whereas the reversed STO/LAO sequence and single-layer structures result in degraded superconducting performance. This difference can be understood in terms of interfacial chemical stability. When LAO is directly deposited on the SAO sacrificial layer, interfacial reactions are promoted, as evidenced by the formation of SrLaAlO$_4$ observed in Films B and C. Such reactions lead to compositional inhomogeneity and are likely responsible for the suppression of $T_\mathrm{c}$ and increased resistivity. In contrast, the insertion of the STO layer between LAO and SAO acts as a chemically stable barrier that suppresses interdiffusion and prevents the formation of secondary phases.

In addition to suppressing interfacial reactions, the STO layer also contributes to maintaining a coherent epitaxial relationship across the heterostructure. The ADF-STEM analysis of Film A reveals sharp interfaces and a cube-on-cube epitaxial alignment throughout the multilayer stack, indicating that strain and lattice continuity are well preserved. Such structural coherence is essential for maintaining the superconducting properties of GdBCO, which are highly sensitive to lattice distortion and local compositional variations. The preservation of a sharp superconducting transition and strong diamagnetic response after lift-off suggests that the optimized LAO/STO bilayer effectively stabilizes the superconducting phase even in the absence of the substrate. These results indicate that the buffer layer design is not only important for epitaxial growth but also plays a crucial role in maintaining superconducting properties in freestanding oxide membranes.

Although electrical transport measurements of fully transferred membranes remain an important subject for future work, the present results already provide consistent evidence that the optimized heterostructure preserves superconducting properties after the lift-off process. In particular, the agreement between the transition temperature before and after lift-off, together with the strong diamagnetic response, supports the robustness of the superconducting state in the freestanding films.
It should be noted that this comparison involves different measurement techniques, namely transport measurements before lift-off and magnetic measurements after lift-off. Therefore, the comparison is intended to provide a qualitative consistency of the superconducting transition rather than a strict quantitative equivalence, and should be interpreted with appropriate caution.

\section{4. CONCLUSION}
In this study, we demonstrated the fabrication of freestanding GdBa$_2$Cu$_3$O$_{7-\delta}$ thin films using a Sr$_3$Al$_2$O$_6$ sacrificial layer and thermal release tape. A LaAlO$_3$/SrTiO$_3$ bilayer buffer was identified as key structural element that maintains epitaxial growth during deposition and preserves this biaxial texture and a superconducting transition temperature of approximately 92\,K after lift-off, comparable to that of the as-grown films. Magnetic susceptibility measurements revealed a sharp superconducting transition and strong diamagnetic response, indicating that superconducting properties are well preserved after the lift-off process. These findings establish a robust heterostructure design strategy for preserving superconducting properties in freestanding oxide membranes and may facilitate their integration into future electronic and flexible devices.

\section*{ASSOCIATED CONTENT}
\textbf{Supporting information}\\
The Supporting Information is available free of charge.\\
AFM image of the TiO$_2$-terminated SrTiO$_3$ substrate; Temperature dependence of resistivity for the films investigated in this study; $T_\mathrm{c}$ as a function Sr content in GdBa$_{2-x}$Sr$_x$Cu$_3$O$_{7-\delta}$; Optical micrographs GdBCO films before and after lift-off and their electro-magnetic properties (PDF)

\section*{AUTHOR INFORMATION}
\textbf{Corresponding author}\\
\textbf{Kazumasa Iida}--\textit{College of Industrial Technology, Nihon University, Narashino, Japan};
\date{Email: iida.kazumasa@nihon-u.ac.jp}; \textbf{ORCID}: 0000-0003-1038-9630\\

\textbf{Authors}\\

\textbf{Kai Walter}--\textit{Institute for Technical Physics, Karlsruhe Institute of Technology, Eggenstein-Leopoldshafen, Germany}; \textbf{ORCID}: 0009-0004-5366-6639\\

\textbf{Takafumi Hatano}--\textit{Department of Materials Physics, Nagoya University, Nagoya, Japan}; \textbf{ORCID}: 0000-0002-6509-5835\\

\textbf{Kose Morinaga}--\textit{College of Industrial Technology, Nihon University, Narashino, Japan}\\

\textbf{Manuela Erbe}--\textit{Institute for Technical Physics, Karlsruhe Institute of Technology, Eggenstein-Leopoldshafen, Germany}; \textbf{ORCID}: 0000-0001-9698-1509\\

\textbf{Hongye Gao}--\textit{The Ultramicroscopy Research Center, Kyushu University, Fukuoka, Japan}\\

\textbf{Satoshi Hata}--\textit{The Ultramicroscopy Research Center, Kyushu University, Fukuoka, Japan; Department of Advanced Materials Science and Engineering, Kyushu University, Kasuga, Fukuoka, Japan; Interdisciplinary Graduate School of Engineering Sciences, Kyushu University, Kasuga, Fukuoka, Japan}\\

\textbf{Jens H\"{a}nisch}--\textit{Institute for Technical Physics, Karlsruhe Institute of Technology, Eggenstein-Leopoldshafen, Germany}; \textbf{ORCID}: 0000-0003-2757-236X\\

\textbf{Notes}\\
The authors declare no competing financial interest.

\section*{ACKNOWLEDGMENTS}
A part of this work was supported by ``Advanced Research Infrastructure for Materials and Nanotechnology in Japan (ARIM)" of the Ministry of Education, Culture, Sports, Science and Technology (MEXT). Proposal Number JPMXP12\,25KU0057. This work was also carried out under the Visiting Researcher's Program of the Institute for Solid State Physics, the University of Tokyo (No. 202411-MCBXG-0008, No. 202505-MCBXG-0098).



\begin{thebibliography}{1}
\bibitem{ref2}
Paskiewicz, D. M.;  Sichel-Tissot, R.; Karapetrova, E.; Stan, L.; Fong, D. D. Single-crystalline SrRuO$_3$ nanomembranes: A platform for flexible oxide electronics. \textit{Nano Lett.} \textbf{2016}, 16(1), 534--542.
\bibitem{ref3}
Bakaul, S.; Serrao, C.; Lee, M.; Yeung, C.; Sarker, A.; Hsu, S-L.; Yadav, A.; Dedon, L.; You, L.; Khan, A.; et al. Single crystal functional oxides on silicon. \textit{Nat. Commun.} \textbf{2016}, 7, 10547.
\bibitem{ref4}
Liu, W.; Wang, H. Flexible oxide epitaxial thin films for wearable electronics: Fabrication, physical properties, and applications. \textit{Journal of Materiomics } \textbf{2020}, 6(2), 385--396.
\bibitem{ref5}
Huang, J.; Chen, W. Flexible strategy of epitaxial oxide thin films. \textit{iScience} \textbf{2022}, 25(10), 105041.
\bibitem{ref6}
Lu, D.; Baek, D.; Hong, S.; Kourkoutis, F. K.; Hikita, Y.; Hwang, H. Y. Synthesis of freestanding single-crystal perovskite films and heterostructures by etching of sacrificial water-soluble layers. \textit{Nat. Mater.} \textbf{2016}, 15, 1255--1260.
\bibitem{ref7}
Harada, T.; Tsukazaki, A. A versatile patterning process based on easily soluble sacrificial bilayers. \textit{AIP Adv.} \textbf{2017}, 7(8), 085011.
\bibitem{ref8}
Takahashi, R; Lippmaa, M. Sacrificial Water-Soluble BaO Layer for Fabricating Free-Standing Piezoelectric Membranes. \textit{ACS Appl. Mater. Inter.} \textbf{2020}, 12(22), 25042--25049.
\bibitem{ref9}
Chen, Z.; Wang, B.; Goodge, B. H.; Lu, D.; Hong, S.; Li, D.; Kourkoutis, L. F.;, Hikita, Y.; Hwang, H. Y. Freestanding crystalline YBa$_2$Cu$_3$O$_{7-x}$ heterostructure membranes. \textit{Phys. Rev. Materials} \textbf{2019}, 3(6), 060801(R).
\bibitem{ref10}
Jia, Z.; Tang, C.; Wu, J.; Li, C.; Xu, W.; Wu, K.; Zhou, D.; Yang, P.; Zeng, S.; Zeng, Z.; et al. Self-passivated freestanding superconducting oxide film for flexible electronics. \textit{Appl. Phys. Rev.} \textbf{2023}, 10(3), 031401.
\bibitem{ref11}
Xie, Z.; Li, Z.; Lu, H.; Wang, Y.; Liu, Y. Etching Sr$_3$Al$_2$O$_6$ sacrificial layer to prepare freestanding GBCO films with high critical current density. \textit{Ceram. Int.} \textbf{2021}, 47(10), 13528--13532.
\bibitem{ref12}
Sandik, S.; Elshalem, B-C.; Azulay, A.; Waisbort, M.; Kohn, A.; Kalisky, B.; Dagan, Y. Freestanding single-crystal superconducting electron-doped cuprate membrane. \textit{Phys. Rev. Materials} \textbf{2025}, 9(2), L021802.
\bibitem{ref13}
Zhang, J.; Lin, T.; Wang, A.; Wang, X.; He, Q.; Ye, H.;  Lu, J.; Wang, Q.; Liang, Z.; Jin, F.; et al. Super-tetragonal Sr$_4$Al$_2$O$_7$ as a sacrificial layer for high-integrity freestanding oxide membranes. \textit{Science} \textbf{2024}, 383, 388--394.
\bibitem{ref14}
Yan, S.; Mao, W.; Sun, W.; Li, Y.; Sun, H.; Yang, J.; Hao, B.; Guo, W.; Nian, L.; Gu, Z.; et al.  Superconductivity in Freestanding Infinite-Layer Nickelate Membranes. \textit{Sci. Adv. Mater.} \textbf{2024}, 36(31), 2402916.
\bibitem{ref15}
Yan, M. F,; Barns, R. L.; O’Bryan, Jr. H. M.; Gallagher, P. K.; Sherwood, R. C.; Jin, S. Water interaction with the superconducting YBa$_2$Cu$_3$O$_7$ phase. \textit{Appl. Phys. Lett.} \textbf{1987}, 51(7), 532--534.
\bibitem{ref16}
Baek, D. J.; Lu. D,; Hikita, Y.; Hwang, H. Y.; Kourkoutis, L. F. Ultrathin Epitaxial Barrier Layer to Avoid Thermally Induced Phase Transformation in Oxide Heterostructures. \textit{ACS Appl. Mater. Interfaces} \textbf{2017}, 9(1), 54--59.
\bibitem{ref17}
Connell, J. G.; Isaac, B. J.; Ekanayake, G. B.; Strachan, D. R.; Seo, S. S. A. Preparation of atomically flat SrTiO$_3$ surfaces using a deionized-water leaching and thermal annealing procedure. \textit{Appl. Phys. Lett.} \textbf{2012}, 101(25), 251607.
\bibitem{ref18}
Gong, L.; Wei, M.; Yu, R.; Ohta, H.; Katayama, T. Significant Suppression of Cracks in Freestanding Perovskite Oxide Flexible Sheets Using a Capping Oxide Layer. \textit{ACS Nano} \textbf{2022}, 16(12), 21013--21019.
\bibitem{ref19}
Asano, H.; Takita, K.; Katoh, H.; Akinaga, H.; Ishigaki, T.; Nishino, M.; Imai, M.; Masuda, K. Crystal Structure of the High $T_\mathrm{c}$ Superconductor LnBa$_2$Cu$_3$O$_{7-\delta}$ (Ln=Sm, Eu and Gd). \textit{Jpn. J. Appl. Phys.} \textbf{1987}, 26, L1410--1412.
\bibitem{ref20}
Seibold, G.; Arpaia, R.; Peng, Y. Y.; Fumagalli, R.; Braicovich, L.; Castro, C. D.; Grilli, M.; Ghiringhelli, G. C.; Caprara, S. Strange metal behaviour from charge density fluctuations in cuprates. \textit{Commun. Phys.} \textbf{2021}, 4, 7.
\bibitem{ref21}
Gunasekaran, R. A.; Hellerbrand, B.; Steger, P. L. Crystal structure, oxygen stoichiometry and superconducting properties of  GdBa$_{2-x}$Sr$_x$Cu$_3$O$_{7-\delta}$ (0.0$\leq x\leq$1.6). \textit{Physica C} \textbf{1996}, 270(1--2), 25--34.
\end{thebibliography}

\newpage

\end{document}